# Lattice resonances under oblique light incidence on nanoparticle array

Viktoriia E. Babicheva

College of Optical Sciences, University of Arizona,
1630 E. University Blvd., P.O. Box 210094, Tucson, AZ 85721

vbab.dtu@gmail.com

**Abstract.** Ultra-thin optical structures, known as metasurfaces, have shown promising light controlling capability at the nanoscale. In this paper, we study their particular case, a periodic array of high-refractive-index nanoparticles with electric and magnetic resonances. The main result of the work is a numerical demonstration that the lattice effect in the periodic arrangement of nanoparticles changes the resonance position even if the resonances are above the diffraction wavelength (Rayleigh anomaly). We show that the disk resonance changes can be achieved not only by varying periods of the array under normal light incidence but also by changing the incident angle.

**Introduction**

Nanoparticles, their arrays, and oligomers are essential building blocks for different optical elements, metasurfaces, and photonic devices. The nanoparticle resonances enable control of light at the nanoscale and efficient scattering, absorption, reflection, transmission, and so on [1]. The interplay of waves scattered by electric and magnetic resonant modes facilitates interesting effects such as directional scattering, Kerker effect [2,3], and strong non-radiative excitations (anapoles) [4]. Various scatterers have been studied in detail: dielectrics [5] and nanoparticles with plasmon- [1,6,7] or phonon-polariton excitations [8], including materials with hyperbolic dispersion [9,10], to name but a few. In the engineered nanostructures, the aim is to achieve the strongest nanoparticle resonances which can be used for efficient light harvesting in photovoltaics [11,12], nanolasers and medical treatment [13], imaging with subwavelength resolution [14], detectors and sensors [15], etc. Recently, high-index dielectric nanoparticles [5,16-28] have been actively explored as building blocks of metasurfaces. Because of the Mie resonance excitations, the high-index nanoparticles facilitate low radiative losses in the structure. At the same time, such designs possess relatively low non-radiative losses in comparison to conventional plasmonic structures.

Periodic arrays of nanoparticles have gained a special attention [29-37] because of extraordinary lattice resonances in the proximity to the wavelength of diffraction (i.e. the wavelength of Rayleigh anomaly, RA). It has been known that the lattice resonances are particularly strong when the RA wavelength is on a red side of the single-particle resonance. The lattice-resonance spectral width depends on its proximity to the single-particle resonance and significantly decreases when the RA wavelength is far away on the red side of the single-particle resonance. In the lattice, effective electric and magnetic dipole moments are defined by the contributions from the lattice of dipoles along the direction of dipole emission. Thus, for instance, for polarization with E-field along the x-axis, the period of the array $D_x$ controls excitation of magnetic dipole lattice resonance and the period $D_y$ controls electric dipole lattice resonance [19,20]. This means the lattice resonances can be controlled independently and brought into overlap by array dimensions tuning [21,22].

In this work, we show that dipole resonances of the array are strongly affected by the lattice, even when they do not overlap with the RA wavelength. We are mainly concerned when the RA wavelength is on the blue side of the single-particle resonance and no additional lattice resonances are excited. We numerically study high-index dielectric disks and identify electric and magnetic Mie resonances in the absorption spectra for different disk diameters, periods of the array, and angles of light incidence. We demonstrate that the resonances change upon variations of the light incidence angle, and we show a relation to the change of array periods. We observe that angle change in transverse magnetic (TM) polarization controls the position of magnetic dipole resonance (MDR) and in transverse electric (TE) polarization, the angle change controls electric dipole resonance (EDR), which is in a good agreement with earlier experimental observations [24]. These results indicate that depending on the initial position of single-particle resonances, the latter can be brought closer and/or into overlap under the proper choice of array periods and angle of light incidence. For the analysis, we intentionally choose disk dipole resonances spectrally separated from each other to show a possibility to control each resonance independently. Alternatively, the resonances can be spectrally close (see experimental work [24]), and one can achieve their overlap with the small changes in the angle.

**Results**

*Normal light incidence*

To start with, we consider a single disk with the height h = 130 nm and different diameters d. The permittivity of the disk is silicon (data from experimental results [38]) and the surrounding material is air with $\varepsilon_o = 1$. We perform numerical modeling with frequency-domain solver of CST Microwave Studio commercial package.

The scattering properties of a single nanoparticle are affected by the interplay of resonances, such as Kerker effect with electric and magnetic dipoles [2], generalized Kerker effect involving higher multipoles [3], or excitation of toroidal mode resulting in non-radiating anapole response [4]. Thus, field enhancement or power absorption in the nanoparticle is the most reliable way to identify resonances. The same holds true for nanoparticle periodic arrangement: one should define resonances in the array absorption spectrum, not in transmittance or reflectance one. Figure 1a shows excitations of the EDR, MDR, and higher multipole resonances for the single disk (inset in Fig. 1a). In this parameter range, the toroidal mode is also excited in the proximity to the EDR [4], but it is relatively weak, and we do not consider it here. From Fig. 1a we observe that at the normal light incidence of a plane wave, the EDR and MDR overlap for d ≈ 350 nm, and the EDR is excited at a larger wavelength for d > 350 nm in contrast to spherical nanoparticle where $\lambda_{MDR} > \lambda_{EDR}$. While multipole resonances appear as maxima in absorption, the total scattering cross-section shows both increased and decreased values around the resonances (Fig. 1b). In particular, the total scattering is increased on the red side of the EDR and decreased on the blue side of the resonance, which is typical for the nanoparticles with such dimensions [4]. Distributions of electric and magnetic fields in xz- and yz-coordinate planes, respectively, confirm the excitation of the EDR and MDR at the moment of corresponding field maximum inside the nanoparticle (insets in Fig. 1b).

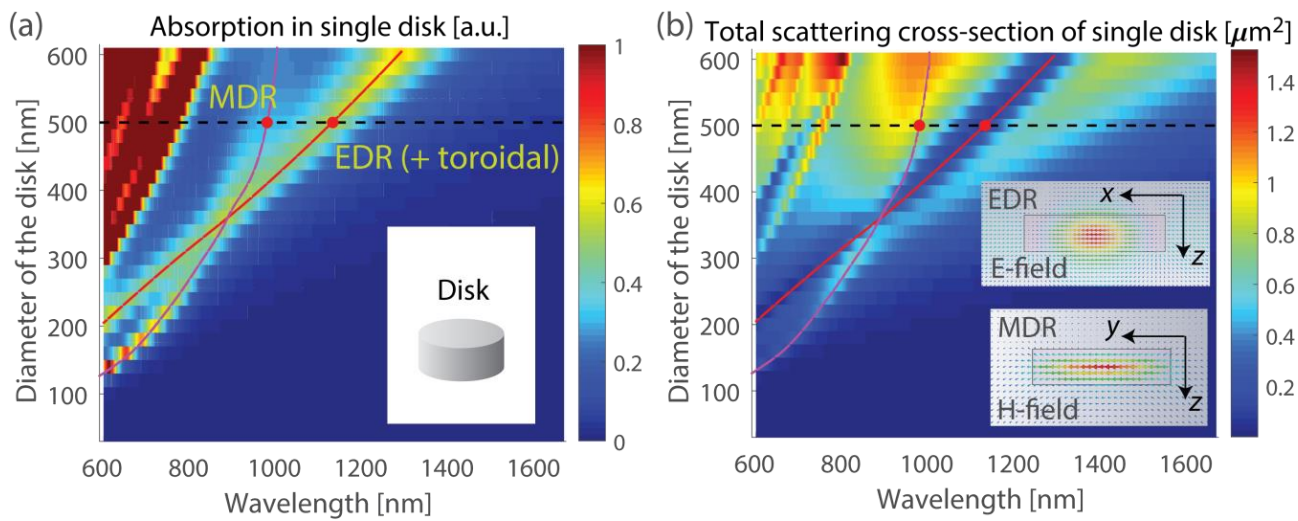

Fig. 1. Absorption and scattering properties of a single silicon disk for various disk diameters under normal light incidence: (a) absorption and (b) total scattering cross-section. The black dotted lines correspond to the diameter d = 500 nm that is chosen for further analysis. The red and magenta lines mark excitation of electric and magnetic resonances, respectively, defined as maxima in absorption. The red dots indicate the wavelength of resonances for d = 500 nm. Inset in (a): schematic of the single disk under consideration. Insets in (b): E-field distribution in the EDR (λ = 1130 nm) and H-field distribution in the MDR (λ = 978 nm) for d = 500 nm. The cross-sections are different as they are in the plane of the corresponding field component (xz-plane for the EDR and yz-plane for the MDR), and the distributions are captured at the moment of corresponding field maximum inside the disk.

Disk diameter d = 500 nm is chosen for the further analysis to ensure the EDR and MDR are spectrally distinguished from each other and the lattice effect can be separately studied for each of the resonance. Now we consider a regular array of such disks with periods $D_x$ and $D_y$ in x- and y-directions, respectively (Fig. 2). The polarization of the incident plane wave is along the x-axis. In this case, period $D_x$ controls the position of the MDR and electric quadrupole (EQ) resonance [35], and period $D_y$ controls the EDR and magnetic quadrupole (MQ) resonance. When both periods are changed simultaneously, both the EDR and MDR experience redshift (Fig. 3). Similar to the total scattering cross-section of the single disk, the reflectance is increased on the red side of the EDR and decreased on the blue side of the resonance. The resulting resonance changes under normal light incidence are summarized in Table 1.

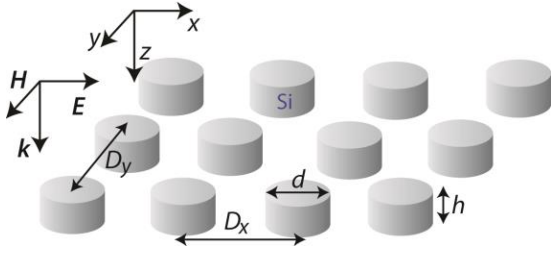

Fig. 2. Schematic view of the silicon disk array in the air. Nanoparticles have diameter d and arranged into the array with the periods $D_x$ and $D_y$. Light incidence **k** is normal to the array plane, and electric field **E** is along the *x*-axis.

Table 1. Lattice effects (LE) under the change of periodicity and angle of light incidence. Under normal incidence, the field is ($E_x$, $H_y$, 0). "LE" (or "No LE") denotes the resonances are affected (or not) by the period changes or angle of incidence for the nanoparticle with scalar polarizability arranged in the infinite array.

| Resonance | $D_x$ | $D_y$ | Oblique TM ($k_x \neq 0 \sim D_x$ change) | Oblique TE ($k_y \neq 0 \sim D_y$ change) |
|---|---|---|---|---|
| ED | No LE | LE | No LE | LE |
| MD | LE | No LE | LE | No LE |
| EQ | LE | No LE | LE | No LE |
| MQ | No LE | LE | No LE | LE |

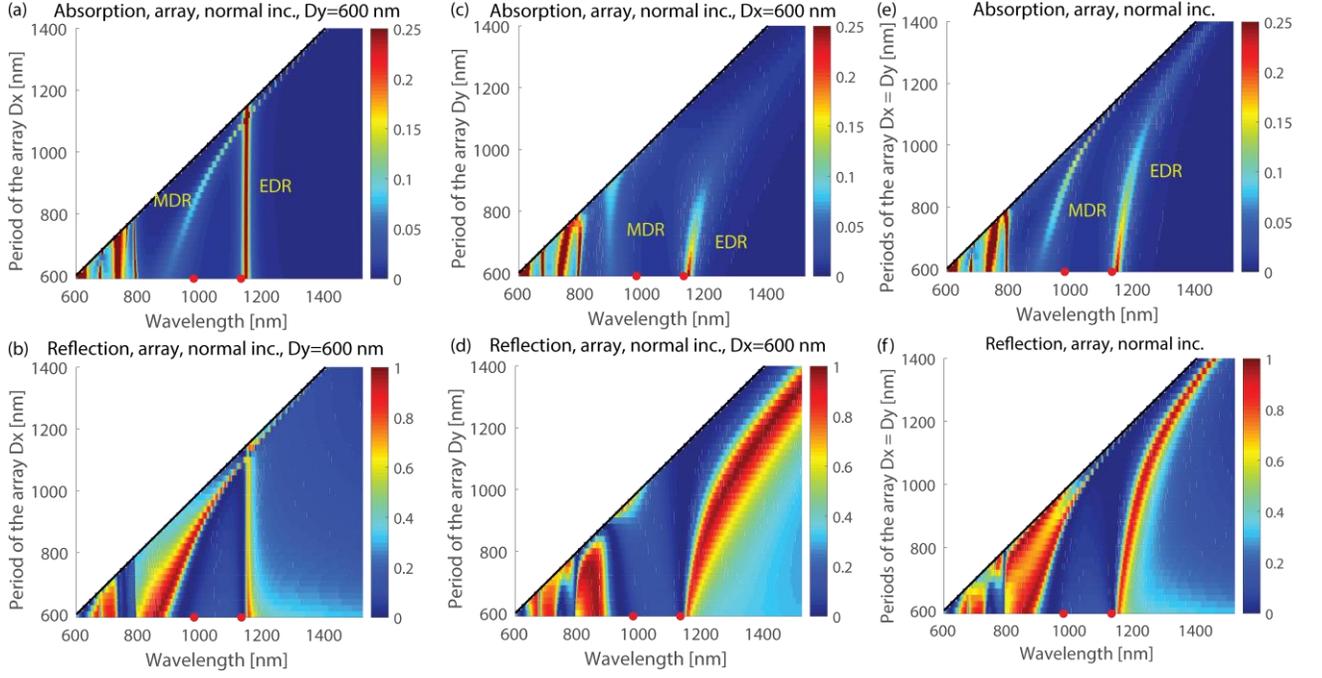

Fig. 3. Electric and magnetic resonances in the arrays of silicon disks under normal light incidence with field components ($E_x$, $H_y$, 0). (a),(c),(e) Absorption and (b),(d),(f) reflection. In panels (a) and (b) period $D_x$ changes, and $D_y$ is fixed. In panels (c) and (d) period $D_y$ changes, and $D_x$ is fixed. In panels (e) and (f) both periods change simultaneously $D_x = D_y$. The red dots mark the position of the resonances of the single disk identified in Fig. 1a. The diffraction wavelength (black lines) is defined as $\lambda_{RA} = D_x$ for (a) and (b), $\lambda_{RA} = D_y$ for (c) and (d), or $\lambda_{RA} = D_x = D_y$ for (e) and (f), and the region below the diffraction is excluded from consideration.

*Oblique light incidence*

Now let us consider an oblique light incidence on the single disk at the different angles θ (Fig. 4a,b). Additional resonances are induced in both TM and TE polarizations: in the former case, field component $E_z \neq 0$ induces extra modes; and in the latter case, $E_x$ remains in the array plane, but the oblique incidence brings an additional component of propagating wavevector $k_y \neq 0$. At the angles θ ≠ 0º the conditions of resonance excitation change. In TM polarization, for 10º < θ < 80º, extra mode appears at the wavelength λ ≈ 950 nm, and for θ > 45º, the resonances correspond to the nanoparticle with inverted aspect ratio and $\lambda_{MDR} > \lambda_{EDR}$. In TE polarization and θ > 30º, extra resonances appear in the range 800-950 nm as well as the MDR at λ ≈ 1520 nm.

Oblique light incidence on the periodic disk array effectively changes the distance between resonant multipoles in the lattice. In this case, the lattice resonance positions shift, and the comparison of array resonances with the single disk resonances reveals the lattice effect. The spectral position of RA can be found from the equation for TM polarization:

$$\left(\frac{2\pi}{D_y}n_y\right)^2 + \left(\frac{2\pi}{D_x}n_x + k_x\right)^2 = k^2, \qquad (1)$$

and for TE polarization:

$$\left(\frac{2\pi}{D_y}n_y + k_y\right)^2 + \left(\frac{2\pi}{D_x}n_x\right)^2 = k^2, \qquad (2)$$

where $n_x$ and $n_y$ are the integers $0, \pm1, \pm2,\ldots$, $k_x$ and $k_y$ are the in-plane components of the incident wave $k_x = k_y = k \sin\theta$, and $k = 2\pi/\lambda_{RA}$ assuming the surrounding medium is air ($\varepsilon_o = 1$). The largest RA wavelength and the only one that corresponds to $\lambda_{RA} > D_{x,y}$ is the case $n_y = 0$, $n_x = -1$, and $\lambda_{RA} = D_x(1+\sin\theta)$ for TM polarization and the case $n_x = 0$, $n_y = -1$, and $\lambda_{RA} = D_y(1+\sin\theta)$ for TE polarization. These diffraction limits are marked in Fig. 4c-f as the black lines.

Importantly, TM polarization corresponds to $k_x \neq 0$ which is an equivalent of $D_x$ change: see Eq. (1). It means that the lattice interaction of the MDR and EQR is affected in TM polarization, and the EDR and MQR are not affected. This conclusion is in a good agreement with results in Fig. 4c,e. In turn, TE polarization corresponds to $k_y \neq 0$ and effective change of $D_y$, affecting lattice interaction of the EDR and MQR, but not affecting the MDR and EQR (Fig. 4d,f). The results of resonance changes under oblique light incidence are summarized in Table 1. The extra mode that appears in TM polarization at the wavelength $\lambda \approx 950$ nm for $10^o < \theta < 80^o$ is not affected by the lattice effect. For TE polarization, the additional resonances and their spectral position changes are defined not only by the lattice effect but also by the change of the excitation conditions of each nanoparticle in the array.

The considered above silicon nanoparticles array is an example of a structure out of realistic material, dispersive and lossy. Such behavior of the EDR and MDR under change of array period and angle of light incidence is observed not only for silicon nanoparticles but also for the nanoparticles with permittivity $\varepsilon = 20 + 0.01i$ (see Figs. 6-8 in Appendix). For the disks with such higher permittivity and lower losses, the multipole resonances are better pronounced and spectrally more separated. Numerical simulations show that the MDR or EDR changes its position under oblique light incidence with TM or TE polarization, respectively, and we expect that these array properties can be observed for a wide range of materials with high refractive index.

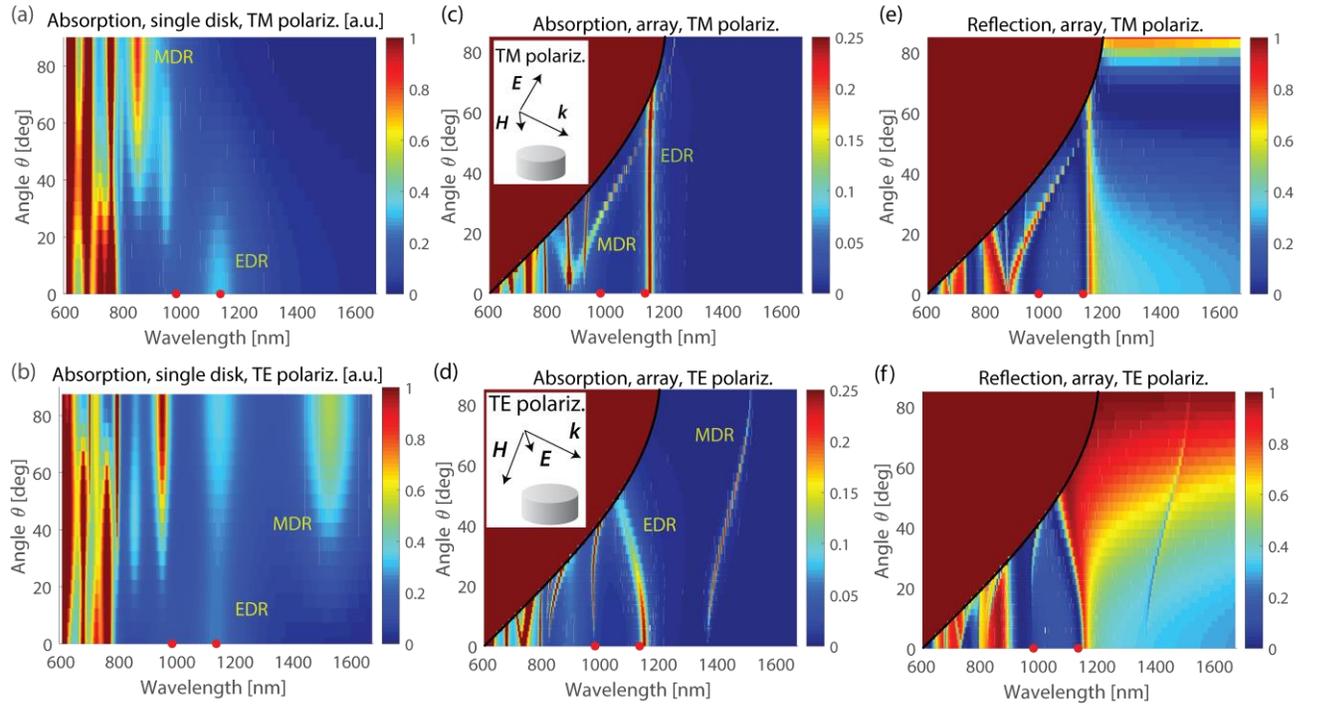

Fig. 4. Electric and magnetic resonances in the single silicon disk and arrays of disks with oblique light incidence. (a) and (b) Absorption in the single disk; (c) and (d) Absorption in the disk array; (e) and (f) Reflection in the array; (a), (c), and (e) are for TM polarization; (b), (d), and (f) are for TE polarization. In (c)-(f), the calculations are performed for a square array with periods $D_x = D_y = 600$ nm. The red dots mark the position of the resonances of the single disk identified in Fig. 1a. The diffraction wavelength (black lines in (c)-(f)) is defined as $\lambda_{RA} = D(1+\sin\theta)$, where $D = D_x$ for TM polarization and $D = D_y$ for TE polarization,

and the region below the diffraction is excluded from consideration (dark red color). Insets in (c) and (d) are the schematics of TM and TE light polarization with respect to the disk, respectively.

Finally, we confirm the effects described above are valid not only for the hypothetical silicon disk array in the air but also for nanoparticles on a substrate and with a top coating. The detailed analytical and numerical analyses of silicon nanoparticle arrays on the substrate have been performed for a variety of substrate indices and nanoparticle shapes (see e.g. [12,25,28]). Substrates with the high refractive index can cause a reflection that significant alternates response of the structure as a whole. However, substrates with the moderate refractive index or uniform coating of nanoparticles do not change the main effects that we discuss in the present work. To illustrate this, we consider the substrate with moderate refractive index n = 1.5, which is close to the index of silicon oxides, glasses, and polymers (Fig. 5a). Although the MDR is significantly weaker than for the case of the uniform environment, it clearly shifts to the larger wavelength upon an increase of the incidence angle θ. At the same time, the EDR remains on the unchanged position. The same is seen for the case of the substrate and top coating with n = 1.5 which effectively mimics a uniform environment around the nanoparticles (Fig. 5b). It is often used in the experimental work [16,24,26] and eliminates the effect of an additional interface and undesirable reflections from it. Thus, as we show by the examples of the substrate with and without top coating, the oblique incidence of light in TM polarization mainly affects the MDR and shifts its position to the larger wavelength moving along the RA wavelength.

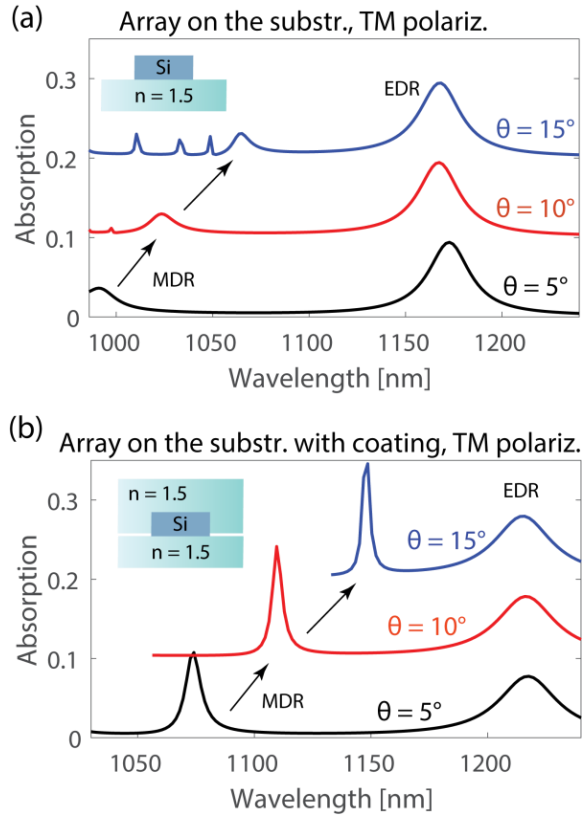

Fig. 5. Effect of the substrate and top coating. (a) Absorption in the silicon disk array in TM polarization in the case of nanoparticles on the substrate with refractive index n = 1.5 (inset shows the basic schematics). Lines for θ = 10º and 15º are shifted by 0.1 with respect to each other for clarity. (b) The same as (a) but with an addition of a top coating of material with the same n = 1.5. The case of the substrate and top coating effectively corresponds to the uniform environment with n = 1.5. The lines for θ = 10º and 15º are interrupted at the RA wavelength. In both cases with and without top coating, the MDR shifts to larger wavelength in agreement with the case without substrate or coating.

**Discussion**

The most remarkable feature of the considered arrays is that even when the RA wavelength is spectrally far away from the disk resonance (100-200 nm in wavelength difference), the lattice arrangement still affects the resonance position. This observation contradicts the misconception that lattice period does not affect resonance position when the resonance wavelength is greater than the diffraction wavelength [16,24,26]. One can see from Fig. 3 that for high-index nanoparticle array, both the EDR and MDR experience a strong redshift upon a change of periods $D_x$ and $D_y$ and the resonances avoid crossing with RA.

For the oblique light incidence on a single disk, we observe that resonances almost do not change their spectral positions. In fact, the resonances may vanish, or new resonances may appear upon the large change of incidence angle. This result is in a good agreement with the previous study of the single silicon disk with a relatively high aspect ratio (height is 220 nm and diameter is 100 nm) and consequently the reverse resonance positions, that is $\lambda_{MDR} > \lambda_{EDR}$ [23].

However, the experimental results in the earlier work [24] have shown that for the closely-packed array of nanoparticles, the EDR and MDR strongly respond to the change of angle of incidence. It has also been reported that the response is qualitatively different in TM and TE polarizations. Motivated by the work [24], here we choose disk with spectrally separated resonances, and we perform analysis of array resonances and their responses to different angles. We show that the main role is played by the arrangement of disks in the array and the lattice effect. Similar to the effect of periods $D_x$ and $D_y$ on the resonance positions under normal light incidence, RA presence under oblique light incidence significantly changes resonance position even though the diffraction wavelength is 100-200 nm spectrally away from the resonance.

In the experiment [24], the disk resonances appear relatively far away from the diffraction. But, as we have just shown, the effect of the lattice becomes pronounced even when resonances do not overlap with the RA wavelength. Increasing the angle of incidence, the RA wavelength is moved and so does the position of either EDR (in TE polarization) or MDR (in TM polarization). Thus, one can explain why it is polarization dependent: oblique incidence in TM polarization effectively changes the distance between magnetic dipoles (see Eq. (1)) and in TE polarization between electric dipoles (see Eq. (2)). In the present work, we have clearly shown that the main effect of resonance shifts in [24] comes from the lattice and the change of RA position and not from moving disk resonances by the angle of excitation.

Multipole calculations in [24] have also revealed an additional electric and magnetic quadrupole resonances excited at the wavelength of dipole resonances and moved along with the latter, but the effect has not been explained. Being orthogonal to the dipole moments (i.e. an integral of the mode overlap is zero), the quadrupole moments do not interact with the dipole moments in the case of the single nanoparticle. As has been shown in our previous work [37], the situation is different for a periodic nanoparticle array, and there, the quadrupole moments are coupled to the dipole moments through the lattice sum. Specifically, the MDR is coupled with the electric quadrupole resonances and by analogy, the EDR is coupled to the magnetic quadrupole resonances. The same relation is observed in [24], and the excitations of quadrupole resonances at the wavelength of dipole resonances confirm the dominating role of the lattice effect in the array.

**Conclusion**

We have studied electric and magnetic resonances in high-index dielectric disks and the lattice effect on the resonance position. First, we analyzed a single silicon disk with various diameters and showed that the absorption spectra should be used to identify resonances rather than the scattering cross-sections. The latter is affected by the interplay of multipole moments and their interference and thus can be misleading. Further, for the identified multipole resonances of the single disk, we showed their control by the lattice dimensions. In contrast to the previous studies, we have shown the lattice effect in the case when the RA wavelength is spectrally far away from the single-particle resonance and on the blue side of it. We have analyzed the oblique light incidence on the disk array and shown that the effects of angle variations on the resonances are similar to the period changes and the resonance position can be controlled accordingly. The MDR changes its position under different angles in TM polarization, and the EDR is responsive to the TE polarization. Our theoretical explanation and numerical results are in a good agreement with the earlier experimental observations of the incidence angle dependence and resonances overlap in the array of silicon disks. This provides an additional degree of freedom in designing metasurfaces and should be taken into account in the analysis of multipole resonances in the periodic arrays.

**Appendix**

Below we show results for the nanoparticles with permittivity $\varepsilon = 20 + 0.01i$ which real part is slightly higher than silicon. Disk height h = 100 nm is adjusted to tune the resonances at the wavelength range the same as for silicon, and the rest of the parameters are kept the same. All effects discussed above are present in this case, and they are better pronounced because of the higher permittivity value and larger separation of the EDR and MDR.

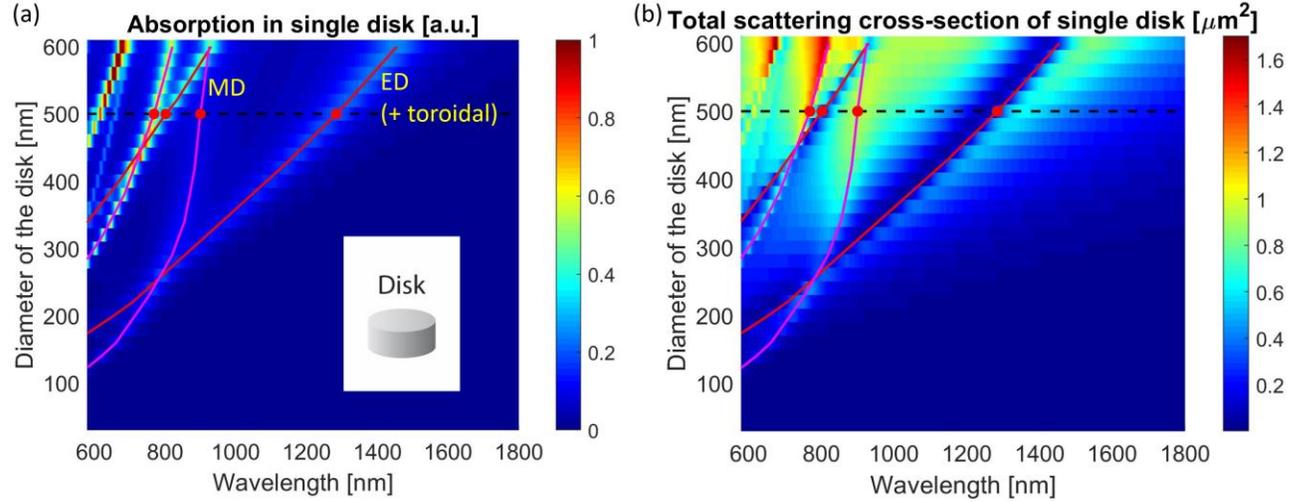

Fig. 6. Absorption and scattering properties of a single disk with permittivity $\varepsilon = 20 + 0.01i$ and h = 100 nm for various disk diameters under normal light incidence: (a) absorption and (b) total scattering cross-section.

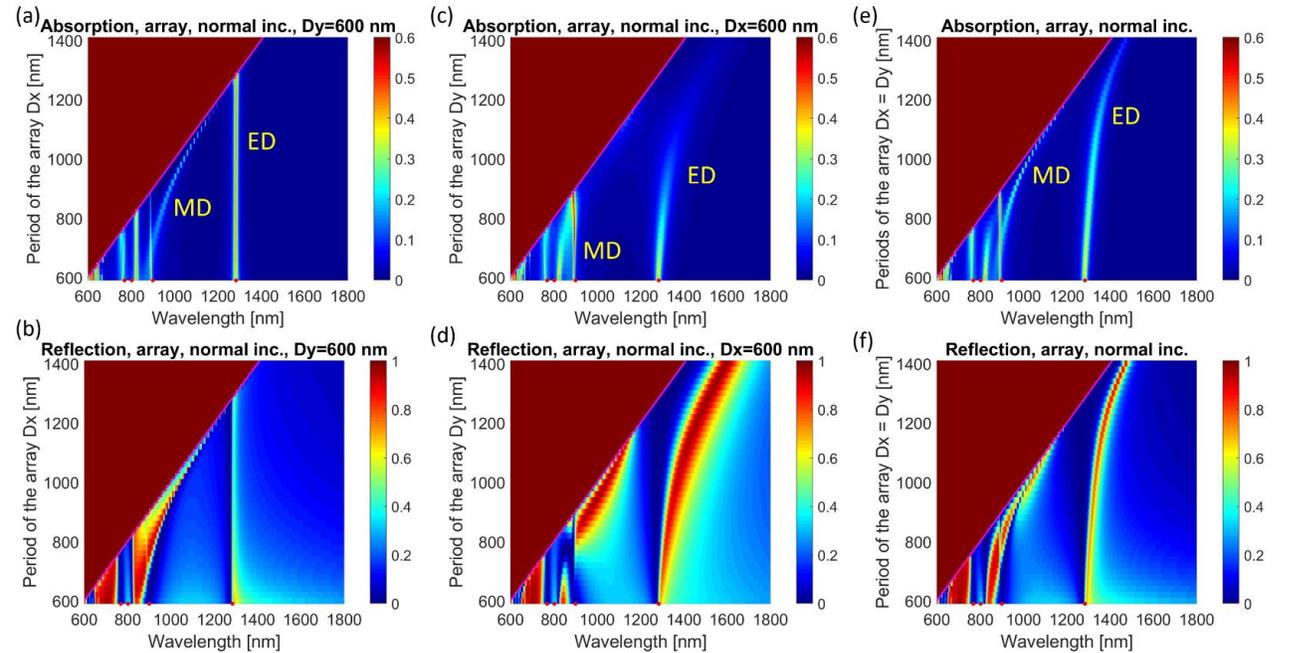

Fig. 7. Electric and magnetic resonances in the arrays of disks with permittivity $\varepsilon = 20 + 0.01i$ and h = 100 nm under normal light incidence with field components ($E_x$, $H_y$, 0). (a),(c),(e) Absorption and (b),(d),(f)

reflection. In panels (a) and (b) period $D_x$ changes, and $D_y$ is fixed. In panels (c) and (d) period $D_y$ changes, and $D_x$ is fixed. In panels (e) and (f) both periods change simultaneously $D_x = D_y$.

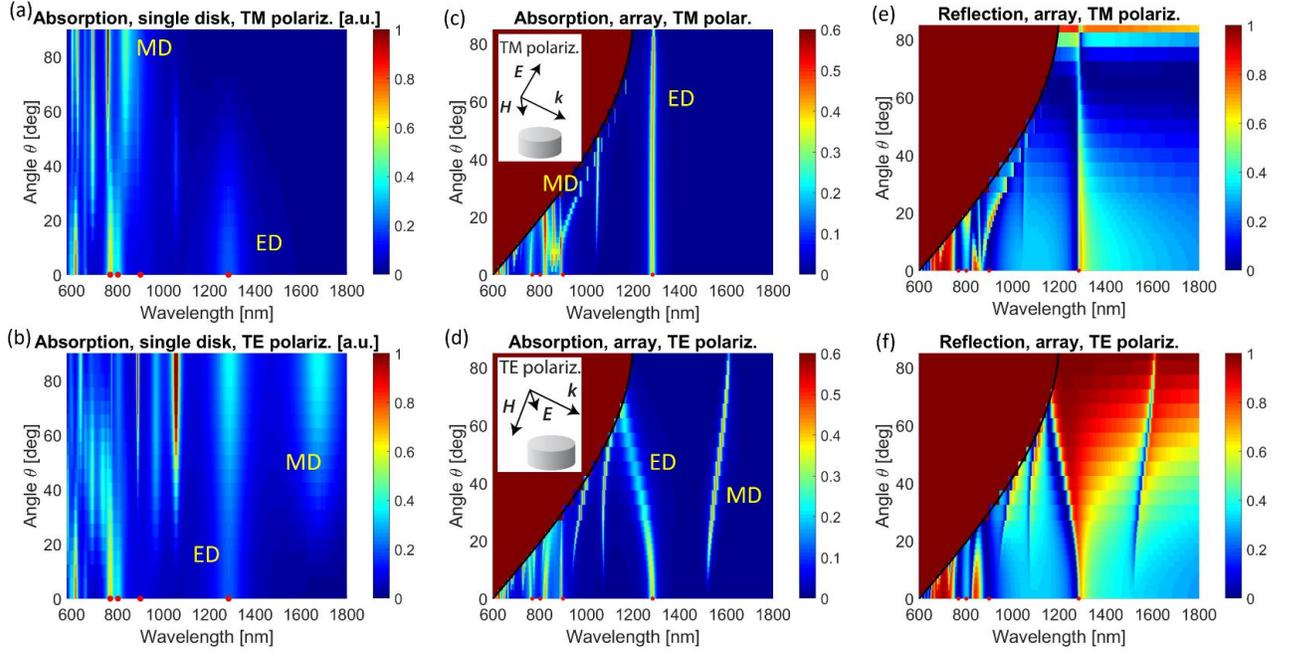

Fig. 8. Electric and magnetic resonances in the single disk and arrays of disks with permittivity $\varepsilon = 20 + 0.01i$ and h = 100 nm for oblique light incidence. (a) and (b) Absorption in single disk; (c) and (d) Absorption in the disk array; (e) and (f) Reflection in the array; (a), (c), and (e) are for TM polarization; (b), (d), and (f) are for TE polarization. For the array, the calculations are performed for a square array with periods $D_x = D_y = 600$ nm. Insets in (c) and (d) are the schematics for TM and TE polarization, respectively. The region below the diffraction is excluded from consideration (dark red color).

**Acknowledgment**

The author thanks Andrey Miroshnichenko for the discussions of the simulation results. This material is based upon work supported by the Air Force Office of Scientific Research under Grant No. FA9550-16-1-0088.

**References**

[1] B. Luk'yanchuk, N. I. Zheludev, S. A. Maier, N. J. Halas, P. Nordlander, H. Giessen, and C. T. Chong: The Fano resonance in plasmonic nanostructures and metamaterials. Nature Materials 9, 707 (2010).

[2] M. Kerker, D. S. Wang, and C. L. Giles: Electromagnetic scattering by magnetic spheres. J. Opt. Soc. Am. 73, 765 (1983).

[3] W. Liu, J. Zhang, B. Lei, H. Ma, W. Xie, and H. Hu: Ultra-directional forward scattering by individual core-shell nanoparticles. Opt. Express 22, 16178 (2014).

[4] A.E. Miroshnichenko, A.B. Evlyukhin, Y.F. Yu, R.M. Bakker, A. Chipouline, A.I. Kuznetsov, B. Luk'yanchuk, B.N. Chichkov, Y.S. Kivshar: Nonradiating anapole modes in dielectric nanoparticles. Nature Communications 6, 8096 (2015).

[5] A.B. Evlyukhin, S.M. Novikov, U. Zywietz, R.L. Eriksen, C. Reinhardt, S.I. Bozhevolnyi, and B.N. Chichkov: Demonstration of Magnetic Dipole Resonances of Dielectric Nanospheres in the Visible Region. Nano Lett. 12, 3749 (2012).

[6] V.E. Babicheva, S.S. Vergeles, P.E. Vorobev, S. Burger, "Localized surface plasmon modes in a system of two interacting metallic cylinders," JOSA B 29, 1263-1269 (2012).

[7] C. Chen, F. Wang, Y. Sheng, and J. Wang: Enhancement transmittance of a metamaterial filter based on local surface plasma resonance. MRS Communications 8, 194 (2018).

[8] W. Streyer, K. Feng, Y. Zhong, A.J. Hoffman, and D. Wasserman: Engineering the Reststrahlen band with hybrid plasmon/phonon excitations. MRS Communications 6, 1 (2016).

[9] V.E. Babicheva: Directional scattering by the hyperbolic-medium antennas and silicon particles. MRS Advances 3, 1913 (2018).

[10] V.E. Babicheva, "Multipole resonances and directional scattering by hyperbolic-media antennas," arxiv.org/abs/1706.07259

[11] S. V. Zhukovsky, V. E. Babicheva, A. V. Uskov, I. E. Protsenko, and A. V. Lavrinenko: Enhanced Electron Photoemission by Collective Lattice Resonances in Plasmonic Nanoparticle-Array Photodetectors and Solar Cells. Plasmonics 9, 283 (2014).

[12] K.V. Baryshnikova, M.I. Petrov, V.E. Babicheva, and P.A. Belov: Plasmonic and silicon spherical nanoparticle antireflective coatings. Scientific Reports 6, 22136 (2016).

[13] W. Zhou, M. Dridi, J. Y. Suh, C. H. Kim, D. T. Co, M. R. Wasielewski, G.C. Schatz, and T.W. Odom: Lasing action in strongly coupled plasmonic nanocavity arrays. Nature Nanotechnology 8, 506 (2013).

[14] V.E. Babicheva, "Surface and edge resonances of phonon-polaritons in scattering-type near-field optical microscopy"

arxiv.org/abs/1709.06274

[15] P. Offermans, M.C. Schaafsma, S.R.K. Rodriguez, Y. Zhang, M. Crego-Calama, S.H. Brongersma, J. Gómez Rivas: Universal scaling of the figure of merit of plasmonic sensors. ACS Nano 5, 5151 (2011).

[16] I. Staude, A.E. Miroshnichenko, M. Decker, N.T. Fofang, S. Liu, E. Gonzales, J. Dominguez, T.S. Luk, D.N. Neshev, I. Brener, Y. Kivshar: Tailoring directional scattering through magnetic and electric resonances in subwavelength silicon nanodisks. ACS Nano 7, 7824 (2013).

[17] W. Zhao, X. Leng, and Y. Jiang, "Fano resonance in all-dielectric binary nanodisk array realizing optical filter with efficient linewidth tuning," Opt. Express 23, 6858-6866 (2015).

[18] Ning An, Kaiyang Wang, Haohan Wei, Qinghai Song, and Shumin Xiao: Fabricating high refractive index titanium dioxide film using electron beam evaporation for all-dielectric metasurfaces. MRS Communications 6, 77 (2016).

[19] A.B. Evlyukhin, C. Reinhardt, A. Seidel, B.S. Luk'yanchuk, and B.N. Chichkov: Optical response features of Si-nanoparticle arrays. Phys. Rev. B 82, 045404 (2010).

[20] V.E. Babicheva and A.B. Evlyukhin: Interplay and coupling of electric and magnetic multipole resonances in plasmonic nanoparticle lattices. MRS Communications 8, 712-717 (2018).

[21] V.E. Babicheva and A.B. Evlyukhin: Resonant lattice Kerker effect in metasurfaces with electric and magnetic optical responses. Laser and Photonics Reviews 11, 1700132 (2017).

[22] Ch.-Y. Yang, J.-H. Yang, Z.-Y. Yang, Z.-X. Zhou, M.-G. Sun, V.E. Babicheva, and K.-P. Chen: Nonradiating Silicon Nanoantenna Metasurfaces as Narrowband Absorbers. ACS Photonics 5, 2596 (2018).

[23] A.B. Evlyukhin, R.L. Eriksen, W. Cheng, J. Beermann, C. Reinhardt, A. Petrov, S. Prorok, M. Eich, B.N. Chichkov, and S.I. Bozhevolnyi: Optical spectroscopy of single Si nanocylinders with magnetic and electric resonances. Scientific Reports 4, 4126 (2014).

[24] D. Arslan, K.E. Chong, A.E. Miroshnichenko, D.Y. Choi, D.N. Neshev, T. Pertsch, Y.S. Kivshar, I. Staude: Angle-selective all-dielectric Huygens' metasurfaces. Journal of Physics D: Applied Physics 50, 434002 (2017).

[25] J. van de Groep and A. Polman: Designing dielectric resonators on substrates: Combining magnetic and electric resonances. Opt. Express 21, 26285-26302 (2013).

[26] M. Decker, I. Staude, M. Falkner, J. Dominguez, D.N. Neshev, I. Brener, T. Pertsch, Y.S. Kivshar: High-efficiency dielectric Huygens' surfaces. Advanced Optical Materials 3, 813 (2015).

[27] V.E. Babicheva and A.B. Evlyukhin, "Resonant suppression of light transmission in high-refractive-index nanoparticle etasurfaces," Optics Letters 43(21), 5186-5189 (2018).

[28] V. Babicheva, M. Petrov, K. Baryshnikova, and P. Belov: Reflection compensation mediated by electric and magnetic resonances of all-dielectric metasurfaces [Invited]. J. Opt. Soc. Am. B 34, D18 (2017).

[29] S. Zou, N. Janel and G. C. Schatz: Silver nanoparticle array structures that produce remarkably narrow plasmon lineshapes. J. Chem. Phys. 120, 10871 (2004).

[30] V. A. Markel: Divergence of dipole sums and the nature of non-Lorentzian exponentially narrow resonances in one-dimensional periodic arrays of nanospheres. J. Phys. B: Atom. Mol. Opt. Phys. 38, L115 (2005).

[31] V.G. Kravets, F. Schedin, and A.N. Grigorenko: Extremely narrow plasmon resonances based on diffraction coupling of localized plasmons in arrays of metallic nanoparticles. Phys. Rev. Lett. 101, 087403 (2008).

[32] B. Auguié and W.L. Barnes: Collective resonances in gold nanoparticle arrays. Phys. Rev. Lett. 101, 143902 (2008).

[33] V.E. Babicheva and J. Moloney, "Lattice effect influence on the electric and magnetic dipole resonance overlap in a disk array," Nanophotonics 7(10), 1663-1668 (2018).

[34] W. Wang, M. Ramezani, A.I. Väkeväinen, P. Törmä, J. Gómez Rivas, T. W. Odom: The rich photonic world of plasmonic nanoparticle arrays. Materials Today 21, 303 (2018).

[35] A.B. Evlyukhin, C. Reinhardt, U. Zywietz, B. Chichkov: Collective resonances in metal nanoparticle arrays with dipole-quadrupole interactions. Phys. Rev. B 85, 245411 (2012).

[36] V.E. Babicheva, "Lattice Kerker effect in the array of hexagonal boron nitride antennas," MRS Advances 3, 2783 (2018).

[37] V.E. Babicheva and A.B. Evlyukhin: Metasurfaces with electric quadrupole and magnetic dipole resonant coupling, ACS Photonics 5, 2022 (2018).

[38] O. S. Heavens: Handbook of Optical Constants of Solids II. J. Mod. Opt. 39(1), 189–189 (1992).